\documentclass[prl,aps,twocolumn,showpacs,floatfix,groupedaddress]{revtex4-1}

\usepackage{graphicx}
\usepackage{dcolumn}
\usepackage{bm}
\usepackage{latexsym}
\usepackage{amsmath}
\usepackage{amssymb}
\usepackage[latin1]{inputenc}
\usepackage[usenames]{color}
\usepackage{wasysym}
\usepackage[usenames,dvipsnames]{xcolor}
\usepackage{multirow}

\usepackage{mathtools}
\usepackage{makecell,tabularx}
\setcellgapes{3pt}

\begin{document}

\title{Air entrainment and granular bubbles produced by an impinging jet of grains into water}

\author{A. M. Cervantes-\'Alvarez{$^1$}, Y. Y. Escobar-Ortega{$^1$}, A. Sauret{$^2$} and F. Pacheco-V\'azquez{$^1$}}

\affiliation{{$^1$}Instituto de F\'isica, Benem\'erita Universidad Aut\'onoma de Puebla, Apartado Postal J-48, Puebla 72570, Mexico \\
{$^2$}Department of Mechanical Engineering, University of California, Santa Barbara, CA, USA}

\date{\today}

\begin{abstract}
A jet of water entering into a pool of the same liquid can generate air entrainment and form bubbles that
rapidly raise to the surface and disintegrate. Here we report the equivalent phenomenon produced by a plunging dry granular jet, so far unexplored. For grains smaller than a critical size, the granular jet entering into the pool produces air entrainment that leads to bubbles formation. The bubbles formed are covered by a shell of grains attached to the bubble air-water interface due to capillary-induced cohesion. 
In contrast to classical air bubbles, these ''granular bubbles" are stable over time because the granular shell prevents coalescence and keeps the air encapsulated either if the bubbles rise to the surface or sink to the bottom of the pool, which is determined by the competition of the buoyant force and the weight of the assembly. Experimentally, we show how the bubble size and volume of entrained air depend on the grain size, liquid properties and jet impact velocity.
\end{abstract}


\maketitle

\begin{figure}[ht!]
\begin{center}
\includegraphics[width=8cm]{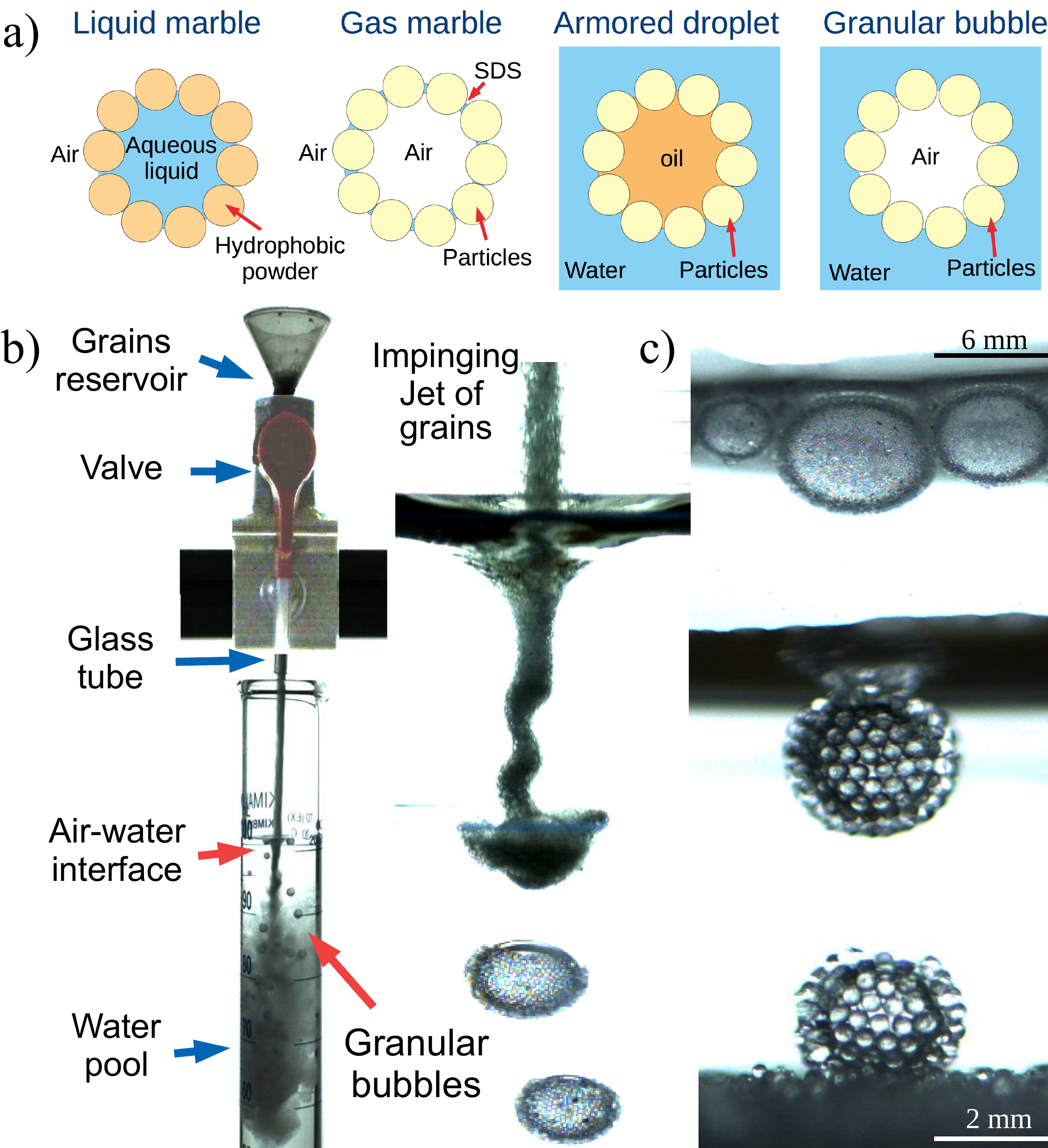} \\
\end{center}
\caption{\small (a) Different types of granular encapsulates depending on the nature of the encapsulated fluid, the cohesive layer of particles and the surrounding fluid. (b) Experimental setup and a picture that shows the formation of granular bubbles. (c) Granular bubbles produced by glass beads of radius $R_g=25$ $\mu$m (top), 125 $\mu$m (middle) and  150 $\mu$m (bottom). 
}
\label{fig1}
\end{figure}

When a granular material is partially saturated with water, liquid bridges form between grains and induce cohesive forces due to the existence of a triple liquid-grain-air interface \cite{Nowak2005,Mitarai2006,Strauch2012}. These forces are responsible for the surprising stability of sandcastles \cite{Pakpour2012} and can also lead to the formation of granular stalagmites when dry grains are poured on a substrate saturated with water\cite{Pacheco2012,Saingier2017}. Capillary forces can also induce the attachment of grains to droplets and generate granular encapsulates, including liquid marbles\cite{Quere2001,Pomeau1999}, gas marbles\cite{Rouyer2017} and colloidal armored droplets\cite{Stone2005,Abkarian2007}, as illustrated in Fig.\ref{fig1}(a). Liquid marbles result from encapsulating aqueous liquid droplets with hydrophobic powder\cite{Quere2001,Pomeau1999,Supakar2017}. Gas marbles are small air pockets encapsulated by a shell of cohesive grains with surfactant solution within an air environment \cite{Rouyer2017}. Armored droplets are structures produced in microfluidic channels that enclose a fluid within a colloidal shell of micrometric-sized beads \cite{Subramaniam2005}. Another mechanism to produce armored droplets is by destabilization of a granular raft floating at the interface between oil and water due to the difference in liquids densities \cite{Abkarian2013,Protiere2017}.  All these granular encapsulates have potential applications including water storage, gas encapsulation, cosmetics, biomedicine, and new aerated materials \cite{Quere2001,Stone2005,Supakar2017,Subramaniam2005,Abkarian2007,Bormashenko2011,Rouyer2017}.

In this article, we investigate the formation of a different granular encapsulate: granular bubbles. We observe that a jet of grains falling into a water pool deforms the air-water interface leading to air entrainment and bubbles formation [Fig.\ref{fig1}(b)]. Although air entrainment induced by a plunging liquid jet into water has been widely investigated \cite{Kigger2012,Prosperetti1992, Bin1993}, the situation with grains remains elusive. Our approach reveals that under certain experimental conditions, bubbles fully covered with grains appear due to the attachment of particles to the bubble air-water interface (see Movie 1\cite{SI}). These granular bubbles can either rise to the surface or sink to the bottom of the pool [Fig. \ref{fig1}(c)] where they remain stable during days. To explore the mechanism of air entrainment, the formation, stability and size of granular bubbles, we systematically varied the grain size, the volume of grains, the jet impact velocity and the physical properties of the liquid.  

\textit{Experimental procedure.-} Before each experiment, a 100 m$\ell$ graduated cylinder is filled with deionized water. A mass $m \in [0.5,\,6.0]\,{\rm g}$ corresponding to a volume $V_{grains} \in [0.2,\,2.3]\,{\rm cm^3}$ of dry glass beads of a given radius $R_g$=25, 50, 75, 125, 150, 200, 225 $\mu$m, and density $\rho_g=$ 2.6 g/cm$^3$ is deposited in a hopper located at a height $h$ above the water level, see Fig. \ref{fig1}(b). When the valve is opened, the grains fall through a vertical glass tube of 4 mm inner diameter and length $h-1$ cm forming a granular jet that reaches the water surface with a velocity $v_j\approx\sqrt{2gh}$. The jet penetration and the subsequent bubble formation are filmed with a high-speed camera Photron UX-100 at 1000 fps. By increasing $h$ from 4 cm up to 61 cm, we varied the jet impact velocity in the range $0.9\,{\rm m.s^{-1}}<v_j<3.5\,{\rm m.s^{-1}}$. The water temperature $T$ was changed in the range of $5^\circ $C $<T<$ $90^\circ $C to systematically decrease the surface tension $\sigma$ in the range $0.075-0.060$ N/m and the dynamic viscosity $\eta$ in the range $1.5-0.32\,{\rm mPa.s}$. These liquid properties were also varied using water with low mole fractions of ethanol, $x=0$, $0.016$ and $0.033$ (see details in \cite{SI}).

\begin{figure}[ht!]
\begin{center}
\includegraphics[width=8cm]{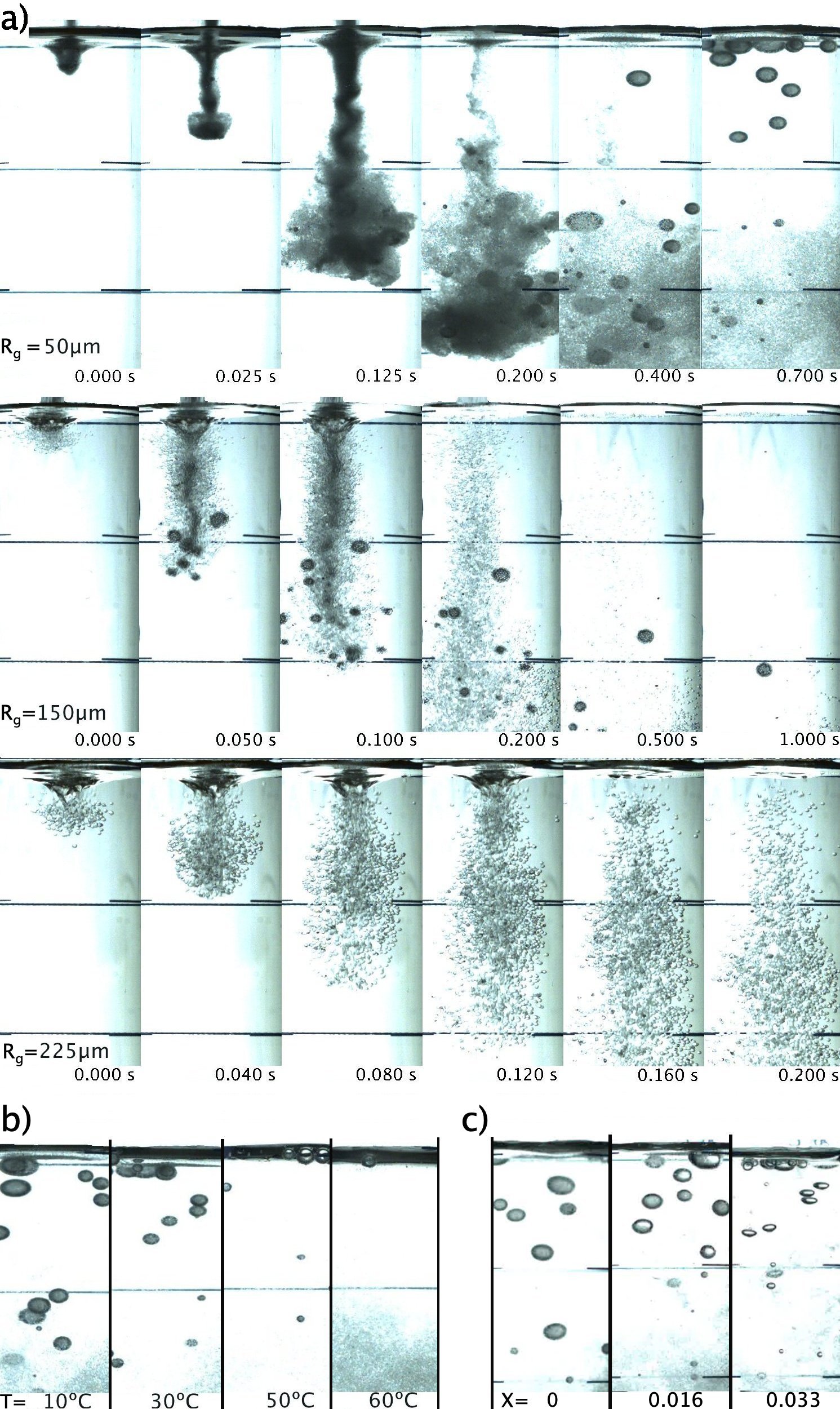}
\includegraphics[width=8cm]{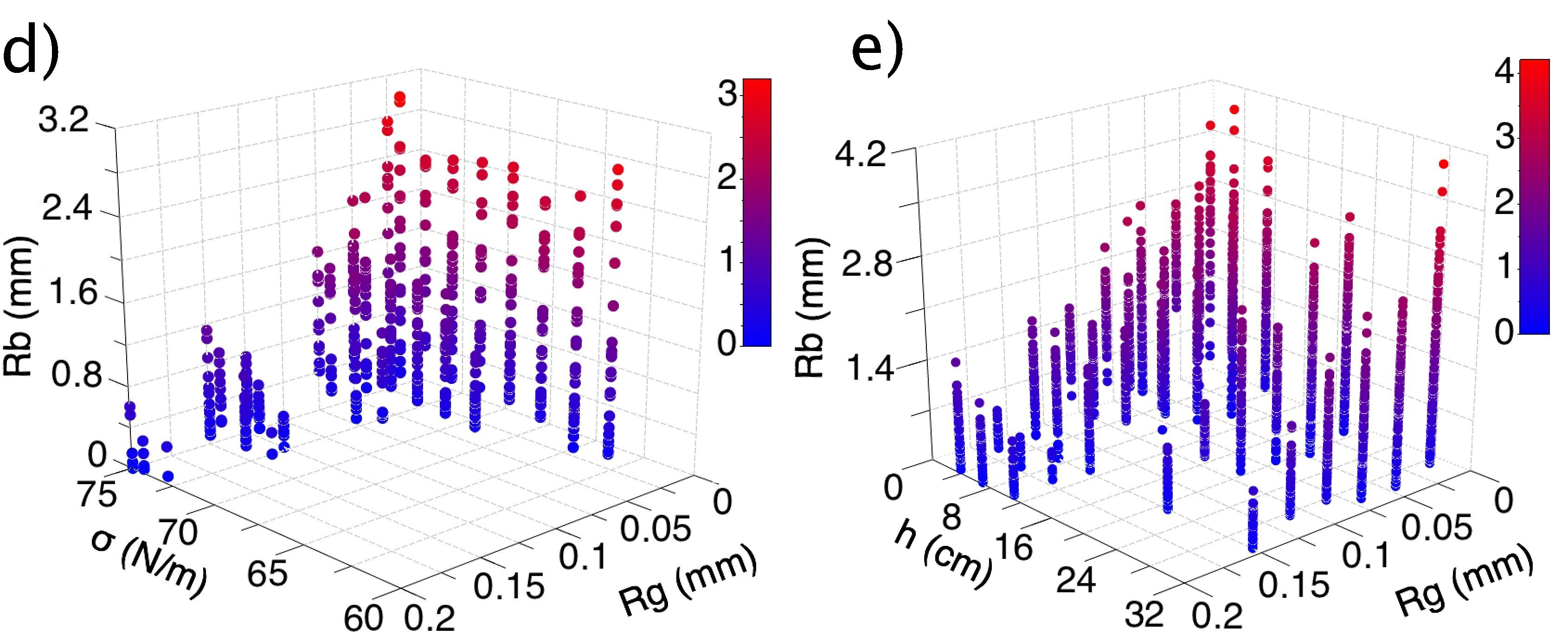}
\end{center}
\caption{\small (a) Effect of the grain radius on the bubble size: for  $R_g=50$ $\mu$m, big granular bubbles rise to the water surface. For $R_g=150$ $\mu$m, the bubbles are smaller and sink to the bottom of the pool. For $R_g = 225$ $\mu$, bubbles are not produced, ($T=20^\circ$C, $h=7.5$ cm)$^*$. (b) Effect of water temperature: the size and number of bubbles decrease as $T$ is increased, ($R_g=50$ $\mu$m, $h=7.5$ cm)$^*$. (c) Bubbles produced in water with three mole fractions of ethanol $x$, ($R_g=50$ $\mu$m, $T=20^\circ$C, $h=7.5$ cm)$^*$. See Movies 2-4 in Ref. \cite{SI}. (d) 3D plot summarizing the data of the bubble radius $R_b$ for different values of $R_g$ and $\sigma$, ($h=7.5$ cm)$^*$. (e) 3D plot of $R_b$ vs $R_g$ and $h$, ($T=20^\circ$C)$^*$. Notation: (constant parameters)$^*$. 
}
\label{fig2}
\end{figure}

\begin{figure*}[ht!]
\begin{center}
\includegraphics[width=17.5cm]{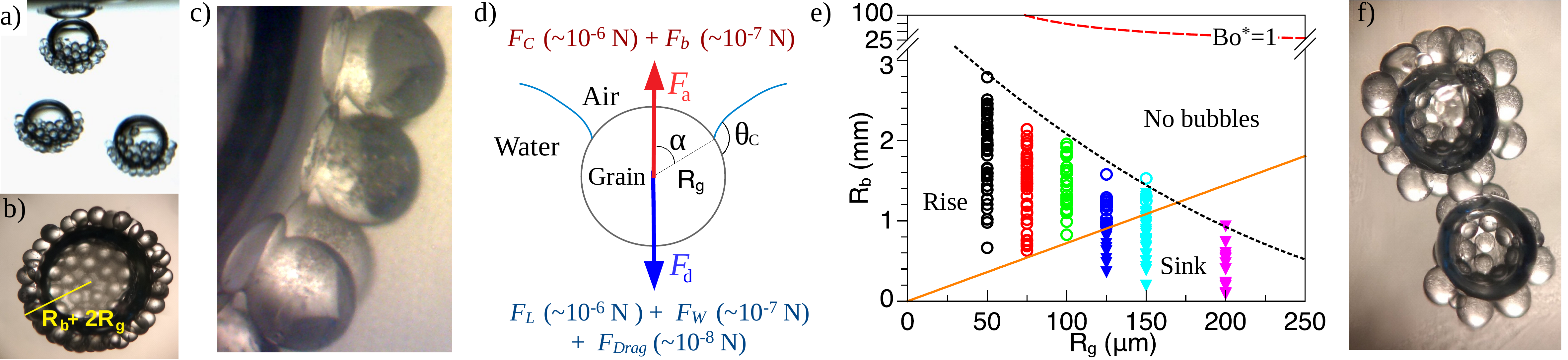}
\end{center}
\caption{\small (a)-(b) Bubbles partially covered with grains allow us to observe that the particles are attached practically outside the bubble air-liquid interface: (a) side view, (b) top view. The total radius of the encapsulate can be approximated as $R_b + 2\,R_g$. (c) Picture taken with a microscope (40X) showing grains of $R_g=150$ $\mu$m attached to the bubble by capillary bridges. (d) Force diagram for a grain attached at the bottom of a bubble (see text). (e) $R_b$ vs. $R_g$ measured at constant $T=20^\circ$C and $h=7.5$ cm. Each empty circle or solid triangle represents a bubble that rises or sinks in the pool, respectively. The orange line represents the condition for rising or sinking of the bubble given by equation \eqref{eq2}. Bubbles are not produced for values above the dotted black line. (f) Image of two granular bubbles separated by a layer of grains that avoids their coalescence.
}
\label{fig3}
\end{figure*}

\paragraph{Results.-} Figure \ref{fig2}(a) shows examples of the jet penetration for a mass $m=0.5$ g of grains of three different radii poured from $h=7.5$ cm into water at $T=20^\circ$C (\textit{i.e}, $\sigma \approx 0.072$ N/m and $\eta=10^{-3}\,{\rm Pa.s}$).  For $R_g \approx 50\mu$m, the jet deforms the air-water interface and penetrates the pool forming a collimated serpentine of grains and air. The destabilization of the granular jet leads to the formation of air bubbles covered by grains that rise to the surface of the pool. As the grain size increases (see the case $R_g=150$ $\mu$m) more grains are scattered across the interface, the serpentine is less pronounced  and the amount of air entrained in the underwater granular jet is less significant. As a result, we observe less granular bubbles, which are also considerably smaller and sink to the bottom of the pool instead of rising. For much larger grains (see the case $R_g=225$ $\mu$m), the jet is scattered, the grains penetrate the pool without entraining air and submerge individually. Consequently, the serpentine is not observed and granular bubbles are not produced during the process.

The size of air bubbles generated by a plunging liquid jet depends on the surface tension $\sigma$ and viscosity $\eta$ of the liquid  \cite{Kigger2012}. To explore the role of these parameters in the formation of granular bubbles, we first varied the water temperature. Figure  \ref{fig2}(b) shows pictures of bubbles produced by grains of $R_g=50$ $\mu$m penetrating in water with different values of $T$. It can be noticed that the granular bubbles are smaller at higher temperatures, i.e. when $\sigma$ and $\eta$ \textit{decrease} \cite{Vazquez1995,Khattab2012,Korson1969}. Above $T \sim 50^\circ$C, granular bubbles are not observed, although some air bubbles may appear the grains do not attach to them.  When binary mixtures of water with low mole fractions $x$ of ethanol are used at $20^\circ$C,  $\sigma$ \textit{decreases} from $0.072$ N/m for $x=0$ (pure water) to $\sim 0.053$ N/m for $x=0.033$, while $\eta$ \textit{increases} from 1 to $1.3$ ${\rm mPa.s}$ \cite{Khattab2012}. Snapshots of these experiments in  Fig.  \ref{fig2}c show that air bubbles appear for $x=0.033$ but the grains do not attach to them. The above observations indicate that the grains stop attaching to the bubbles when the surface tension decreases, regardless of the variation of the liquid viscosity.

Figures \ref{fig2}(d)-(e) summarize the data obtained from our experiments. The bubble size was measured using ImageJ and plotted as a function of $R_g$, $\sigma(T)$ and $h$. Each point represents one bubble. Note that the biggest bubbles of radius $R_b \sim 3-4$ mm form for the smallest grain size. When $R_g$ increases, the number of bubbles and their size decreases, and their formation is only possible above a certain value of surface tension, see Fig. \ref{fig2}(d). The dependence of $R_b$ with $h$ does not show a monotonous trend [Fig. \ref{fig2}(e)].  In the following, we discuss the particle attachment process. Then, we rationalize the condition for a granular bubble to rise or sink inside the pool, and finally, we report the volume of entrained air as a function of the poured mass, grain size and impact velocity.

\begin{figure*}[ht!]
\begin{center}
\includegraphics[width=18cm]{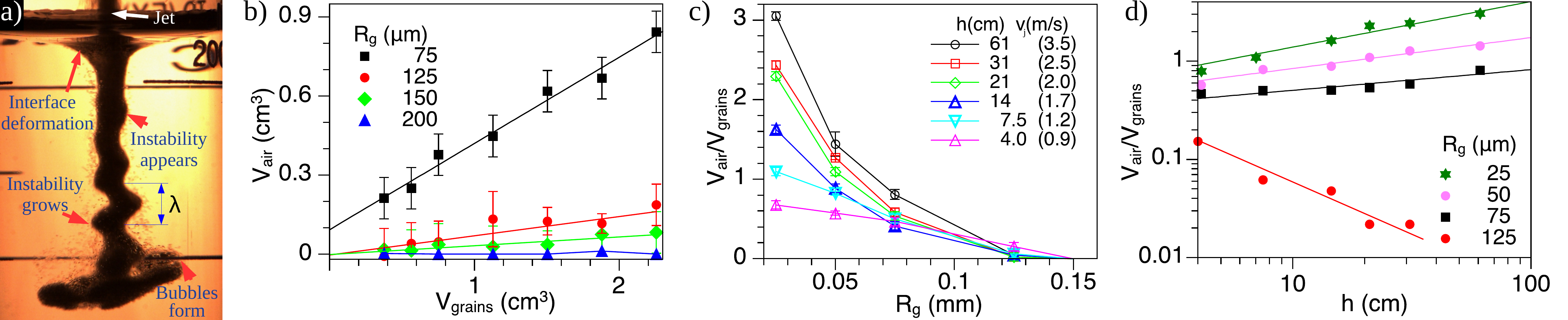}
\end{center}
\caption{\small Air entrainment: (a) Picture of a granular jet entering into water showing the instabilities that originate granular bubbles. (b) $V_{air}$ vs $V_{grains}$ for different grain size and $h=7.5$ cm. (c) $V_{air}/V_{grains}$ as a function of $R_g$ for $m=0.5$ g and different impact heights (the equivalent impact velocity is indicated).  (d) Log-log plot of  $V_{air}/V_{grains}$ as a function of $h$ for $m=0.5$ g. Solid lines correspond to power-laws of the form $V_{air}/V_{grains}=A h^p$, with $p=0.47\pm0.08$ (green line), $p=0.41\pm0.04$ (pink line), $p=0.3\pm0.1$ (black line) and $p=-0.9 \pm0.1$ (red line).   
}
\label{fig4}
\end{figure*}

\paragraph{I. Particle attachment and bubble stability.-}
In most of our experiments, the bubbles are entirely covered with particles, but for grains of radius $R_g = 150$ $\mu$m, we observe some bubbles only partially covered with grains that accumulate towards the bottom of the encapsulate due to gravity, see Figs. \ref{fig3}(a)-(b). A closer view shown in Fig. \ref{fig3}(c) allows us to observe that each grain is practically attached outside of the air bubble by a capillary bridge, but no liquid bridges are visible between particles. Thus, a grain remains attached to the bubble if the attaching forces $F_a$ balance the detaching forces $F_d$. Figure \ref{fig3}(d) shows a force diagram for a particle attached to the bottom of a bubble \cite{Note1}. In that position, the attaching forces are the capillary force $F_c = 2\pi \sigma R_g \sin\alpha \sin( \theta_c-\alpha)$ and the buoyancy $F_b=\pi R_g^3 \rho_l g (2-3\cos\alpha -\cos^3 \alpha)/3$. The detaching forces are the pressure force $F_L=2 \sigma  \pi  R_g^2 \sin^2\alpha /R_b $ due to the Laplace pressure difference $\Delta P = 2\sigma/R_b$, the weight of the particle $F_w=4\pi R_g^3 \rho_g g/3$, and the drag force $F_{Drag}=6\pi \eta R_g v_j$. In these expressions, $\rho_l$ is the water density, $\alpha$ is the half-central angle indicated in the diagram, and $\theta_c$ is the contact angle for the triple air-water-grain interface \cite{Note2}. From pictures as the one shown in Fig. \ref{fig3}(c), we measured values of $\alpha\sim 30^{\circ}$ and $\theta_c\sim 70^{\circ}$ used to estimate the orders of magnitude of the attaching and detaching forces indicated in Fig. \ref{fig3}d.  Then, following Refs. \cite{Wang2016-1,Wang2016-2}, we determined the maximum expected bubble size $R_b^{max}$ able to keep particles of radius $R_g$ at the bubble interface by taking the modified Bond number $Bo^*= F_d/ F_a=1$. This condition is plotted as a dashed red line in Fig. \ref{fig3}(e). Remarkably, bubble radii $R_b\sim \mathcal{O}(30-100$ mm) obtained from the force analysis are one order of magnitude larger than the experimental values of $R_b$ vs $R_g$, of $\mathcal{O}(3 mm)$, represented by color symbols in Fig. \ref{fig3}(e). This indicates that the maximum bubble size is not given by the above stability condition.

According to Ref. \cite{Bin1993}, a review about air-entrainment produced by impinging water jets, there is a general agreement that the air bubbles in the rising bubble region have maximum diameters of 3-4 mm independently of the jet velocity and nozzle diameter. Something similar is found in our case: $R_b^{max} \sim 3-4$ mm. Even if the grains could attach to considerably larger bubbles, as it was estimated above, such bubbles are not produced during the process. In our experiments, the dependence of the maximum bubble size on the grain size is well described by the expression $R_b = R_b^{max}(1-R_g/R_g^*)^2$ represented by the dotted black line in Fig. \ref{fig3}e , where $R_g^*$ is the maximum radius of grains that produce bubbles and it decreases with $\sigma$. The value of $R_b^{max}$ is of the same order of magnitude than the capillary length $\lambda_c=\sqrt{\sigma/(\rho\,g)}=2.7\,{\rm mm}$. Classical air bubbles larger than $\lambda_c$ may fragment after their formation and reorganize in a smaller distribution of bubbles following a coalescence-aggregation scenario \cite{Villermaux2007}. Such scenario is not observed with granular bubbles due to the existence of a grain monolayer that prevents coalescence between adjacent bubbles  [Fig. \ref{fig3}(f)] or between the bubble and the interface [Fig. \ref{fig1}(c)]. All these results explain why granular bubbles have such a long stability and can survive for several days.

\paragraph{II. Rising or Sinking condition.-}
Let us consider a bubble of radius $R_b$ totally covered by a monolayer of $N$ glass beads of radius $R_g$ occupying a surface fraction $\phi= N \pi R_g^2/ 4\pi (R_b + R_g)^2$. Then, the number of particles attached to the bubble is approximately:
$N\approx 4 \phi (R_b+R_g)^2/R_g^2$. Being $V_g$ the volume of one bead, the armoured bubble of volume $V_b = (4/3) \pi R_b^3 + NV_g$ will rise (resp. sink) in the liquid if the buoyant force $B$ is larger (resp. smaller) than the weight of the encapsulate $W$ given by weight of the air bubble $W_{air}$ plus the weight of the attached grains $W_{grains}$. Since $W = W_{air} + W_{grains} \approx W_{grains} = N \frac{4}{3}\pi R_g^3 \rho_g g,$ and $B= \rho_l g V_b = \frac{4}{3} \pi \rho_l g \left(R_b^3 +4\phi(R_b+R_g)^2R_g\right),$  the condition for the equilibrium $B=W$ is satisfied by:
\begin{equation}
R_b^3-4\phi \frac{\rho_g-\rho_l}{\rho_l}(R_b+R_g)^2 R_g=0.
\label{eq2}
\end{equation}

From equation \eqref{eq2}, we can obtain $R_b$ as a function of $R_g$ assuming a monolayer of identical spheres in hexagonal close packing $\phi_{hcp}$ covering the bubble as observed in Fig. \ref{fig1}(c).  Using $\phi_{hcp}=0.84$ for a two-dimensional hexagonal lattice, the condition given by Eq. \eqref{eq2} is plotted together with the experimental data in Fig. \ref{fig3}(e). Note that the measured radius of bubbles rising to the surface (open circles) and sinking to the bottom of the pool (solid triangles) are separated in two zones well-predicted by the proposed model (orange line). \\

\paragraph{III. Air entrainment by the granular jet.-}
In Ref. \cite{Prado2011}, it was found that dry glass beads flowing out from a funnel form a smooth and collimated jet if the grain size $d$ and the funnel outlet size $D$ satisfy the inequality $D/d>15$; otherwise the flow of grains is discrete like with distinguishable grains falling in air under the action of gravity. If we do the calculations for our experiments considering that $D= 4$ mm and $d=2R_g$, one finds $D/2R_g \approx 40$ for $R_g=50 {\rm \mu m}$, while  $D/2R_g \approx 8$ for $R_g=225$ $\mu$m. Therefore, our observations shown in Fig.\ref{fig2}(a)  of a collimated jet for the smallest grains and a dispersed jet for the largest ones are in agreement with Ref. \cite{Prado2011}. Furthermore, when the collimated jet penetrates at high speed into the water pool, the jet-liquid interaction generates instabilities that trigger the buckling of the slender granular structure, see Fig. \ref{fig4}(a). The jet adopts a serpentine shape that thins and breaks into bubbles when the instability has a wavelength $\lambda \approx 5$mm. 
Since air entrapment was only observed when the jet produces the collimated serpentine, the above results suggest that the volume of air entrained by individual grains is negligible and that the air entrained by the jet is mainly in its interstitial space, i.e. proportional to $V_{grains}$. To quantify the volume of air $V_{air}$ entrained by a volume of grains $V_{grains}$, we measured the total change of the liquid volume $\Delta V = V_{air}+ V_{grains}$ inside the graduated cylinder. Figure \ref{fig4}(b) shows that $V_{air}$ is indeed linearly proportional to $V_{grains}$. Moreover, the slope decreases with the grain size and is practically zero for $R_g=200$ $\mu$m (when the jet enters dispersed). Accordingly, Fig. \ref{fig4}(c) shows that the ratio $V_{air}/V_{grains}$ decreases with $R_g$ and becomes negligible for $R_g > 150$ $\mu$m. In addition, Fig. \ref{fig4}(d) reveals that $V_{air}/V_{grains}$ depends on the impact height $h$ approximately as a power law of the form $V_{air}/V_{grains} = Ah^p$, where the exponent $p$ decreases with the grain size in the studied range. 

In the above analysis, an appropriate scaling in terms of dimensionless numbers (for instance, the capillary number or the Weber number) is not straightforward. Even for the classical problem of a water jet entering into a pool of the same liquid, after decades of research \cite{Bin1993,Tran2013,Aristoff2009}, there are no simple analytical expressions to determine the amount of air entrainment given the large number of parameters involved in the problem (see the recent review by Kiger \& Duncan \cite{Kigger2012}). Here, the system becomes considerably more complex due to the addition of the grains dynamics (which is complex even for the case of a single sinking sphere \cite{Vella2015}). 
Since air entrainment is relevant in industry \cite{Kigger2012}, for instance, in controlled aeration for decarbonation, oxidation, stripping and bacteria control\cite{GE2012}, or in the natural generation of tsunamis by landslides \cite{Viroulet2014}, but also because the mechanism reported here can be used to stabilize bubbles and foams, our research could inspire the community in investigating further the air entrainment conditions and bubbles generated by plunging granular jets. 

\begin{acknowledgments}
This research was supported by CONACYT Mexico and the VIEP-BUAP project 2019. AMCA and YYEO thank the PhD scholarships provided by CONACYT Mexico.
\end{acknowledgments}

\bibliographystyle{apsrev}

$^*$ Corresponding author: fpacheco@ifuap.buap.mx

\bibliography{Biblio-BB}

\end{document}